# Transport properties of YBa$_2$Cu$_3$O$_{7-\delta}$ thin films near the critical state with no applied magnetic field

P.Bernstein,  J.F.Hamet and Y.Thimont

*CRISMAT-ENSICAEN (UMR-CNRS 6508) - Université de Caen-Basse Normandie, Boulevard du Maréchal Juin F14050 Caen cedex, France*

e-mail : pierre.bernstein@ensicaen.fr

**Abstract :** Transport measurements carried out on twinned YBa$_2$Cu$_3$O$_{7-\delta}$ films are compared to the predictions of a previously proposed model suggesting that the vortices move along the films twin boundaries that behave as rows of Josephson weak links [P.Bernstein and J.F.Hamet, J.Appl.Phys.**95** (2004) 2569]. The obtained results suggest that, except if the films are very thin, the twin boundaries consist of superimposed rows of weak links with mean height, $\overline{d}_s$, whose mean length along the TBs is an universal function of T/T$_c$, the reduced temperature. This conclusion yields a general expression for the critical surface current density of the films as a function of T/T$_c$ and of the number of superimposed weak links rows, while the critical current density depends on $\overline{d}_s$. A comparison of the measurements reported by various authors shows that the nature of the substrate and the growth technique have both a strong effect on $\overline{d}_s$. The existence of superimposed weak links rows is attributed to extended defects generated by Y$_2$O$_3$ inclusions.

**I. INTRODUCTION**

The role of extended planar defects in the physics of $YBa_2Cu_3O_{7-\delta}$ (YBCO) films, i.e. low angle grain boundaries and twin boundaries (TBs), has been discussed for a long time [1, 2, 3]. Many authors have pointed out that since the width of these planar defects is in the range of $\xi_{ab}(T)$, the superconducting coherence length in the a-b planes of YBCO, some type of Josephson behavior could be expected. Gurevich and Colley [4,5] consider that in presence of a network of planar defects parallel to the flux lines, the vortices whose cores are located on these defects are Abrikosov vortices with Josephson cores. Mezzetti *et al.* consider that the boundary planes behave as long Josephson junctions whose coupling energy is modulated by defects [6].

In the case of films epitaxially grown on $SrTiO_3$ substrates, electron microscopy observations have shown that the TBs can stretch over several micrometers [7]. In addition, magnetic susceptibility measurements suggest that the TBs planes act as grooves channeling the vortices parallel to the TBs and that vortex pinning occurs at the TBs intersections [8]. The vortex channeling effect of the TBs was previously pointed out by various types of measurements [9,10]. Then, in this work we consider that the planar defects that are important for the transport properties of these films, when no external magnetic field is applied, are the twin boundaries. At a smaller scale, HREM (High Resolution Electron Microscopy) observations have revealed the non uniformity of the twin boundaries at the scale of a few inter-atomic distances [11,12]. The defects correspond to tiny atomic displacements, variation in atoms coordination or local vacancies. The average size of this disorder is three cells in the a-b planes for the most coherent TBs. In addition, the existence of TBs is associated to a lattice rotation which is not 90° but around 89.1°, which implies a lattice distortion and strain fields at the TBs locations that change locally the TBs width. Since the TBs width is in the range of the coherence length, the modulation of this width by disordered areas is expected to have a strong effect on the TBs Josephson behavior.

In previous works [13,14], we have suggested that the main effect of the modulation of the TBs width is that the separation between the superconducting banks of the TBs can be locally large enough with respect to $\xi_{ab}(T)$ to cause a disruption of the tunneling current. This is expected to result in the splitting of the TBs into rows of Josephson weak links (see Fig.1). This model has yielded predictions on the value of $I_J$, the current flowing across each weak link in the critical state when no magnetic field is applied and on the value of the vortex pinning energy.

In this paper, we develop this model. In section II we remind of some of its aspects, relevant for the present work. In section III, we determine the mean length of the weak links included in the TBs of thin YBCO micro-bridges with width w≈10μm deposited on SrTiO$_3$ substrates. We determine $I_J(T)$ and we show that the obtained values are in the vicinity of those predicted by the model. We suggest that films with a thickness larger than d ≈10 nm include superimposed rows of weak links along the TBs. Correcting accordingly the measured weak links mean lengths we show that the corrected values collapse into an unique curve. We confirm these results by deriving and verifying experimentally an expression for $J_{cr}^S\left(\frac{T}{T_c}\right) = J_{cr}d$, the critical surface current density of the films ($J_{cr}$ is their critical current density). From measurement of the critical current density in YBCO films carried out by other authors, we show that the expressions of $I_J(T)$ and $J_{cr}^S\left(\frac{T}{T_c}\right)$ are valid for films deposited on various substrates by different methods, but that they don't fit the results in some cases. Section IV is devoted to a discussion, in particular on the possible origin of the existence of superimposed weak links rows in thick films.

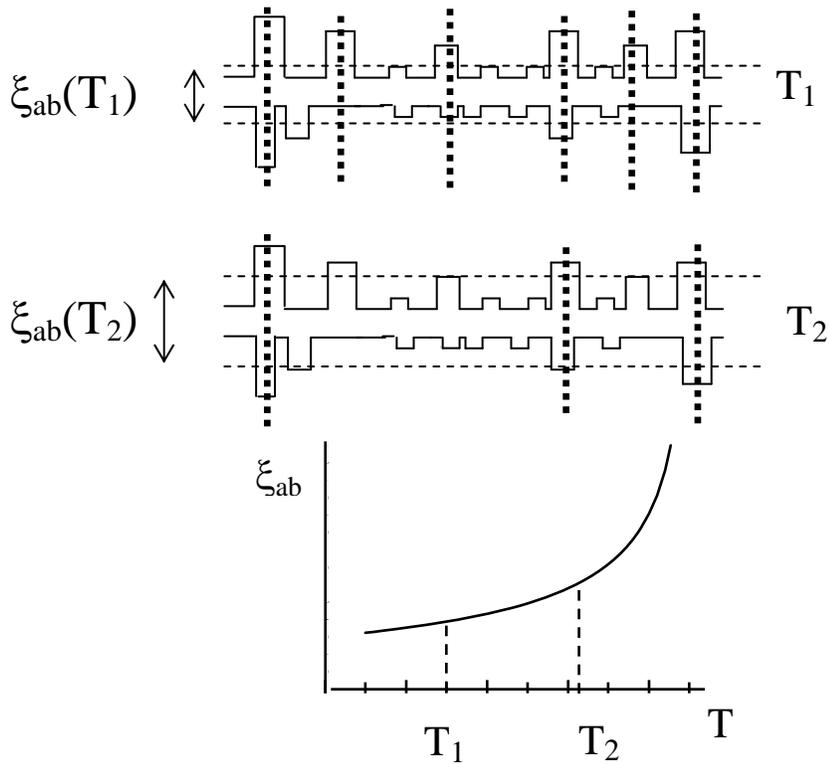

Fig.1 : Schematic representation of a twin boundary section at two different temperatures ($T_1<T_2$) and sketch of the dependence on temperature of the coherence length, $\xi_{ab}$. The rectangular peaks represent disordered areas (or defects) lying along the TB that change locally its width. The tunneling current is disrupted wherever the effective separation between the superconducting banks is much larger than the coherence length. As a result the TBs consist of rows of weak links bounded by large disordered areas (at the atomic scale). The thick dashed lines represent the limits of these weak links. Since at temperature $T_1$ the coherence length is shorter than at temperature $T_2$, more defects or disordered areas are large enough to disrupt the superconducting current at $T_1$ than at $T_2$. Consequently, the weaks links are shorter at $T_1$ than at $T_2$.

**II- THE CRITICAL STATE OF YBCO FILMS**

In this section we summarize some features of the model we have developed in Refs.[13, 14] for YBCO films in the critical state with no applied magnetic field. The model is valid for epitaxial films whose thickness is in the range of or smaller than $2\lambda_{ab}$ ($\lambda_{ab}$ is the superconducting penetration depth in the a-b planes of YBCO) and where twinning is the main structural defect.

When bias current I flows in a YBCO film, the current and the flux first enter its peripheral part, while the central area remains flux and current free for $I<I_{cr}$ [15]. Magneto-optical observations have shown that the TBs are the first areas of the samples penetrated by the magnetic flux [16]. As a consequence, we assume that the TBs accommodate vortices in the region penetrated by the flux. The area of the current and flux free central region decreases as the current increases. When this region disappears, vortices and anti-vortices fill up the TBs and the film is in the critical state ($I=I_{cr}$). For $I>I_{cr}$, as confirmed by scanning SQUID microscopy, additional vortices and anti-vortices enter the film [17]. Gurevich has suggested that the longitudinal pinning force for vortices localized on planar defects can be much smaller than that of vortices located in the bulk of the films and that this result is independent of the details of the pinning potential [4]. We assume that for $I \geq I_{cr}$, as an effect of thermal activation, in spite of pinning at the TBs intersections [8], the vortices and anti-vortices located on the TBs move in direction of the middle of the film where they annihilate each other. In the critical state, all the weak links along the TBs are expected to carry current, except those at the vortex cores. The disorder due to the existence of defects in the weak links and considerations on the vortices and TBs energies suggest that in the critical state the weak link energy is equal to $k_BT$. Then, all the weak links, whatever their length, carry the same net current that is equal to

$$I_J = \frac{2\pi k_B T}{\phi_o} \qquad (1)$$

where $\phi_o$ is the flux quantum. Since the density of superconducting pairs decreases as the temperature increases, Eq.(1) is supposed to be valid for all the weak links below an upper temperature, $T_{up}$, only [13,14]. Above this temperature, the TBs can include weak links, whose maximum superconducting current is smaller than the value given by Eq.(1), which carry no current in the critical state of the film. The value of $T_{up}$ can be estimated from the dependence on temperature of the vortex pinning energy. Experimental verifications of Eq.(1) are reported in section III.

## III. SUPERCONDUCTING PROPERTIES OF THE INVESTIGATED FILMS

In this section we briefly describe the samples and how they were patterned and measured. Then, we detail how the mean length of the weak links was determined from the current-voltage curves (CVCs) and we report the obtained results. The values obtained for bridges thicker than d≈10nm being not consistent with the physics of superconductivity, we suggest that these films include superimposed rows of weak links. This suggestion yields an empirical expression for the mean length of the weak links as a function of the reduced temperature, as well as an expression for $J_{cr}^S\left(\frac{T}{T_c}\right)$. We compare the experimental results to the model predictions. In the final sub-section, from critical current density measurements carried out by various authors, we calculate the corresponding $J_{cr}^{S}\left(\frac{T}{T_c}\right)$ and $I_J(T)$ values, that we compare to those provided by the model.

### A. Experimental details

#### 1. Films

We report the measurements carried out on micro-bridges patterned from four YBCO films named F1, F2, F3 and F4. The fabrication process of the samples and some properties of micro-bridges patterned from films F1, F2 and F3 were reported in Ref.[13]. Briefly, YBCO films were laser ablated on (100) SrTiO3 substrates and in-situ capped with a 20 nm thick $La_4BaCu_5O_{13}$ over-layer to avert damage during patterning [18] and a gold layer to facilitate contact. Micro-bridges with various dimensions were patterned from the films by conventional photolithography and chemical etching. A schematic representation of the patterning masks was reported in Ref.[13]. In this work, we present the results obtained with micro-bridges patterned with a 10 μm wide mask and, in addition, with a micro-bridge patterned from film F2 with a 20μm wide mask. In the figures, the micro-bridges patterned from film F2 with the 10 μm wide and the 20 μm wide masks will be referred to as F2-10μm and F2-20μm, respectively. The micro-bridges exact dimensions as well as their thickness, the value of $T_{up}$ and their offset critical temperature are reported in Table 1. As seen in this table, the micro-bridges width is different from that of the masks. This results from chemical etching. The making process is also probably responsible for the difference in the critical temperatures of the micro-bridges patterned from film F2.

TABLE I. Thickness d, width w, length L, offset critical temperature $T_c$ of the investigated microbridges; $T_{up}$ is the temperature above which Eq.(1) is not valid. It is estimated from the dependence on temperature of the vortex pinning energy [13] ; $\nu$ is the assumed number of flux quanta carried by the vortices and Z that of the superimposed rows of weak links along the TBs ; $\overline{d}_s = \dfrac{d}{Z}$ is the mean height of the weak links. Two micro-bridges with different dimensions were patterned from film F2.

| Film | d (nm) | w (μm) | L (μm) | $T_c$ (K) | $T_{up}$ (K) | $\nu$ | Z | $\overline{d}_s$ (nm) |
|---|---|---|---|---|---|---|---|---|
| F1 | ≅10 | 13.5 | 18.8 | 86.1 | 70 | 2 | 1 | ≅10 |
| F2 | ≅8 | 13.5; 23 | 18.8; 36.3 | 85; 83 | 70; 74.6 | 2; 2 | 1; 1 | ≅8 |
| F3 | 40 | 12.5 | 18.5 | 88.5 | 85.9 | 1 | 7 | 5.7 |
| F4 | 20 | 12.6 | 18.6 | 88.3 | 79.8 | 2 | 3 | 6.7 |

**2. Measurement method**

The current-voltage characteristics of the micro-bridges were measured in a large temperature range by the four-point method. The maximum measurement current had to be kept low with respect to the critical current to make sure that, except at the TBs intersections, vortex-vortex interactions were negligible. However, it is possible to fit accurately the CVCs (see section III-A-3) only if the measurements extend over at least two decades in voltage. Then, the measurements were carried out with a current below 1.4 $I_{cr}$, for which the above conditions were roughly fulfilled. The measurements were carried out with a F=1kHz triangular current and, as a result, a part of the voltage was inductive and frequency dependent. The inductive contribution resulted in uniform offset voltages equal to -$V_{off}$(F) and +$V_{off}$(F) for increasing and decreasing currents, respectively. These contributions were subtracted from the measured values. We have never observed inductive non-linear behaviour and we have verified that no changes were visible in the CVCs after this correction when changing the frequency of the measurements between 100Hz and 5kHz.

**3. Determination of the mean length of the weak links along the TBs in the micro-bridges**

Since the vortex motion in twinned YBCO films is thermally activated in the critical state, the current-voltage curves are generally described with expressions derived from the Kim-Anderson equation

$$V = V_o \exp\left(\frac{-U(I)}{k_B T}\right) \qquad (2).$$

In Eq.(2), V is the voltage measured at the sample terminals, U(I) the vortex activation energy or pinning energy and $V_o$ a constant. The expression of U(I) has been widely discussed in the literature, especially when magnetic field B is applied, but it is generally assumed that it takes the form of a power law with exponent µ. From a general point of view, µ depends on the current range, magnetic field, and temperature for collective creep and vortex glass models. In the case of collective pinning, Feigel'man *et al.*[19] have claimed that µ depends on the size of the moving vortex bundles and is equal to 1/7 for isolated vortices. Zeldov *et al.*[20] have suggested that, due to the shape of the pinning wells, U(I) has a logarithmic form that yields the power law

$V(I) = V_o \left(\dfrac{I}{I_{cr}}\right)^n$ with $n = \dfrac{U(T,B)}{k_B T}$ for the CVCs. Here, we consider isolated vortices in the

vicinity of the critical state with no applied field and little interactions, except at the TBs intersections [14]. We suppose that U(I)=U(0)-W(I) where work W is carried out by the current when a vortex line with length $\ell$ moves over a distance equal to the mean length of a weak link, $\bar{\delta}$. Work W is written as

$$W = J\Phi\ell\bar{\delta} \qquad (3).$$

In Eq.(3), J is the current density in the film and $\Phi$ the magnetic flux carried by a moving vortex that takes the form $\Phi = \nu\phi_o$, where $\nu$ is an integer. Measurements reported in Refs.[13,14] have shown that $\nu$ can be larger than unity, although in a classical description, the vortices carry one flux quantum only, in order to maximize the superconducting-normal interfaces. This is probably due to the fact that the vortices do not settle anywhere in the film but are constrained to move along the TBs. The expression

$$V = R_A (I - I_{cr}) \sinh\left(\frac{I}{I_o}\right) \qquad (4a)$$

with

$$\frac{I}{I_o} = \frac{W}{k_B T} \qquad (4b)$$

where $R_A$, $I_{cr}$ and $I_o$ are the fitting parameters, was proposed in Ref.[13] for fitting the CVCs. In Eq.(4a) the hyperbolic sine function is introduced to take into account the possibility of backward vortex jumps, as is usually done when dealing with thermal activation. Otherwise, Eq.(2) takes into account the vortices created by an external source and neglects the auto-induced vortices, while Eq.(4a) takes into account the auto-induced vortices only. Since V is proportional to the density of additional vortices entering the micro-bridge when $I>I_{cr}$ [17], it is proportional to $[I-I_{cr}]$. It is possible to determine the pinning energy [see Eqs.(6), (7), (10) and (11) in Ref.[13] and the value of ν for each film from the fitting parameters. The ν values obtained for the investigated micro-bridges are reported in Table I. Supposing in a first step that the length of the moving vortex lines is equal to the film thickness, from Eqs.(3) and (4b) the mean length of the weak links along the TBs can be written as

$$\bar{\delta}_{exp} = \frac{wk_B T}{I_o \Phi} \qquad (5).$$

**B. Results**

1. **Critical current density**

Fig.2 shows the critical current density of the investigated samples as a function of $T/T_c$. These values were determined from the values obtained for the critical current when fitting the CVCs with Eq.(4a). They are very similar, but $J_{cr}\left(\frac{T}{T_c}\right)$ is slightly larger for the 20nm and the 40nm films than for the films with d≤10nm.

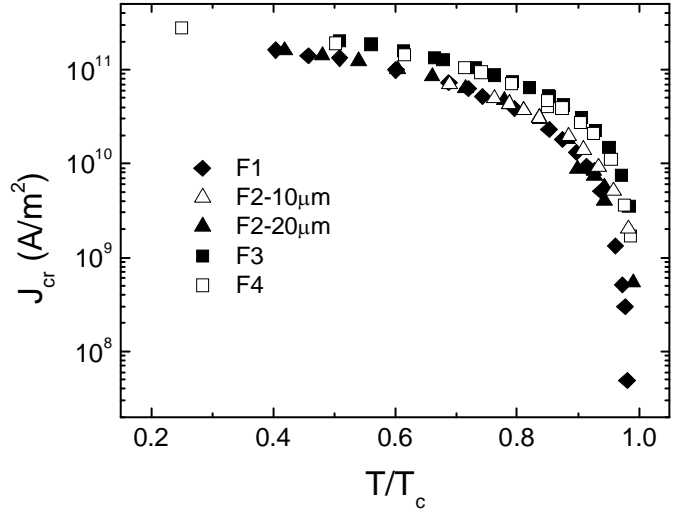

Fig.2 : Critical current density of the micro-bridges patterned from films F1 to F4, as functions of the reduced temperature $\frac{T}{T_c}$.

## 2. Mean weak link length

Fig.3 compares $\overline{\delta}_{exp}$ (T/T$_c$), as calculated for the four micro-bridges, to $\xi_{ab}$(T/T$_c$), as given by the Ginzburg-Landau expression, taking $\xi_{ab}$(0)=1.41nm [21]. The $\overline{\delta}_{exp}$ values of the micro-bridges patterned from films F1 and F2, that have a thickness in the 10nm range, are larger than $\xi_{ab}$ in a large domain of temperature. The $\overline{\delta}_{exp}$ values of the micro-bridges patterned from films F3 and F4, that are 40nm and 20nm thick respectively, are smaller than $\xi_{ab}$, except near T$_c$ [22]. As a result, for these two films $\overline{\delta}_{exp}$ cannot be regarded as the mean length of their weak links, since this length can't be less than the mean distance between the electrons of a Cooper pair. We suggest that in very thin films there is a single row of weak links whose height is equal to the film thickness and with mean length $\bar{\delta} = \bar{\delta}_{exp}$. At the opposite, the TBs of thick films include Z superimposed rows of weak links with mean height $\bar{d}_s = \dfrac{d}{Z}$. The possible origin of the existence of superimposed weak links rows along the TBs is discussed in section IV. While the vortex lines are expected to be rigid in films with TBs including a single row of weak links, they are expected to consist of Z segments if the TBs include superimposed rows. Then, the vortices behave as pancake vortices and work W is not the work corresponding to the motion of a vortex line with length d over distance $\bar{\delta}_{exp}$, but that corresponding to the motion of a vortex line section with mean length $\ell = \bar{d}_s$ over distance $\bar{\delta} = Z\bar{\delta}_{exp}$. Fig.4 shows the $\bar{\delta}$ values computed with the $\bar{\delta}_{exp}$ (T/T$_c$) values reported in Fig.3 and the Z values of Table 1. For all the micro-bridges, the $\bar{\delta}$ (T/T$_c$) values collapse into a single curve that can be fitted with the empirical expression

$$\bar{\delta} = \delta_o (1 - T/T_c)^{-\frac{3}{2}} \qquad (6)$$

where $\delta_o = 0.55\,\text{nm}$. We observe that $\bar{\delta}$ (T/T$_c$) is an increasing function of the temperature. This behavior is consistent with the suggestion that more and more disordered areas are effective for disrupting the Josephson current as the temperature decreases, reducing the mean length of the weak links (see Fig.1). To confirm that the above suggestions are relevant, in the following sections we examine the weak link current and the surface current density of the micro-bridges in the critical state.

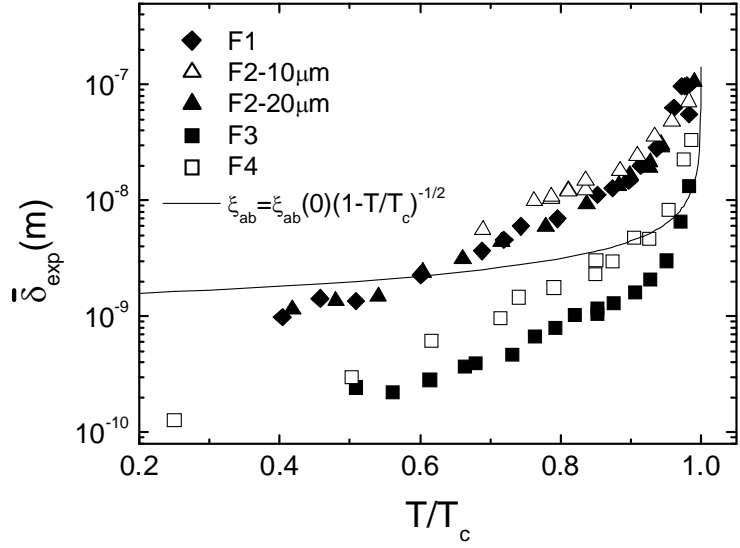

Fig.3 : Length $\bar{\delta}_{exp}$ (T/T$_c$), as computed with Eq.(5) from the current-voltage curves of the micro-bridges patterned from films F1 to F4. The dotted line shows the coherence length in the a-b plane of YBCO, $\xi_{ab}$(T/T$_c$), as given by the Ginzburg-Landau expression with $\xi_{ab}$(0)=1.41nm.

## 3. Weak link current in the critical state

In the critical state, the mean current flowing across a weak link takes the form

$$I_J = \frac{I_{cr}}{\bar{N}} \qquad (7a),$$

where

$$\bar{N} = Z\frac{w}{\bar{\delta}} = \frac{w}{\bar{\delta}_{exp}} \qquad (7b)$$

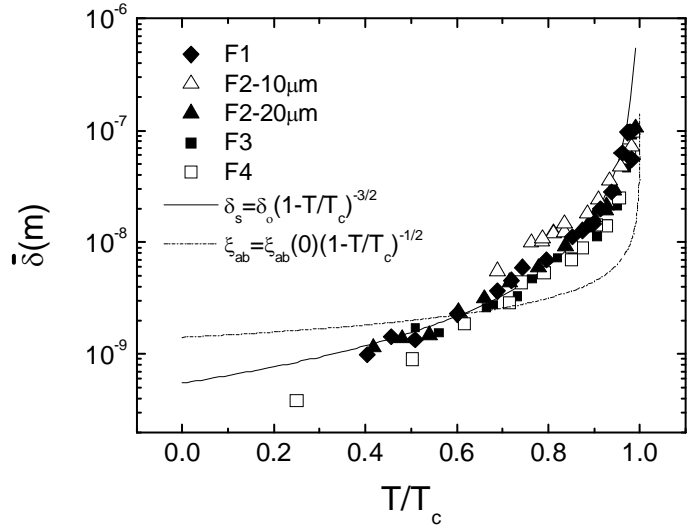

Fig.4 : Mean weak link length $\bar{\delta}$ (T/T$_c$) of the micro-bridges patterned from films F1 to F4 as obtained from the $\bar{\delta}_{exp}$ (T/T$_c$) values reported in Fig.3 and the Z values in Table I. The solid line is obtained with Eq.(6) and the dashed line shows $\xi_{ab}$(T/T$_c$).

is the mean number of weak links included in a TB. Then, $I_J$ can be written as

$$I_J = J_{cr} d\bar{\delta}_{exp} = J_{cr} \frac{d\bar{\delta}}{Z} \qquad (7c).$$

Fig.5 shows the $I_J$ values computed with Eq.(7c) from the $J_{cr}$ and the $\bar{\delta}_{exp}$ values that are reported respectively in Figs.2 and 3. We stress out that no hypothesis on the value of Z is made when estimating $I_J$ with these quantities. For all the films $I_J(T)$ is zero at T=T$_c$ and increases as the temperature decreases down to temperature $T_{up}$. At lower temperatures, the $I_J$ values cluster around those computed with Eq.(1). The low $I_J$ values computed with Eq.(7c) above $T_{up}$ are consistent with the suggestion that in this domain of temperature the number of current carrying weak links is lower than $\bar{N}$ in the critical state. Below $T_{up}$, the good agreement between the experimental $I_J$ and those computed with Eq.(1) confirms that the weak link energy in the critical state is equal to $k_B T$.

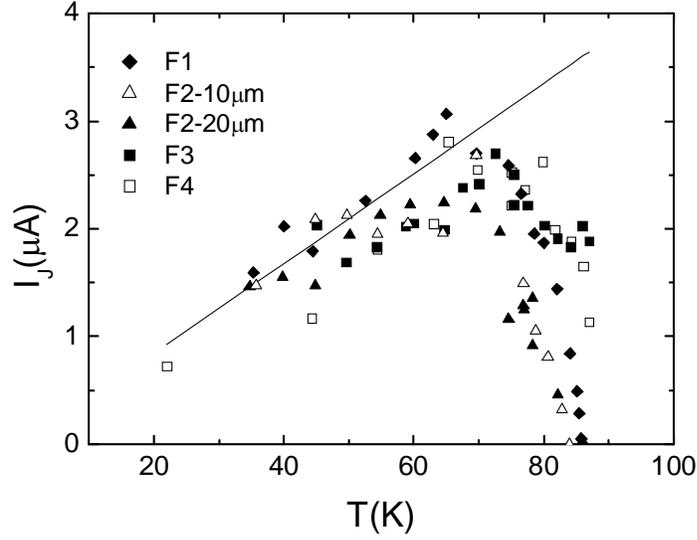

Fig.5 : Current $I_J$ flowing across a weak link of the micro-bridges patterned from films F1 to F4 in the critical state as a function of the temperature. The $I_J$ values are computed with Eq.(7c) and the $J_{cr}$ and $\bar{\delta}_{exp}$ values reported in Figs.2 and 3, respectively. The solid line shows $I_J(T)$ as computed with Eq.(1).

## 4. Surface current density in the critical state

Since each weak link carries current $I_J$ in the critical state, the critical surface current density of a micro-bridge including Z rows of weak links can be written as

$$J_{cr}^S = Z\frac{I_J}{\bar{\delta}} = Z\frac{2\pi k_B T}{\phi_o \delta_o}\left(1 - \frac{T}{T_c}\right)^{\frac{3}{2}} \qquad (8).$$

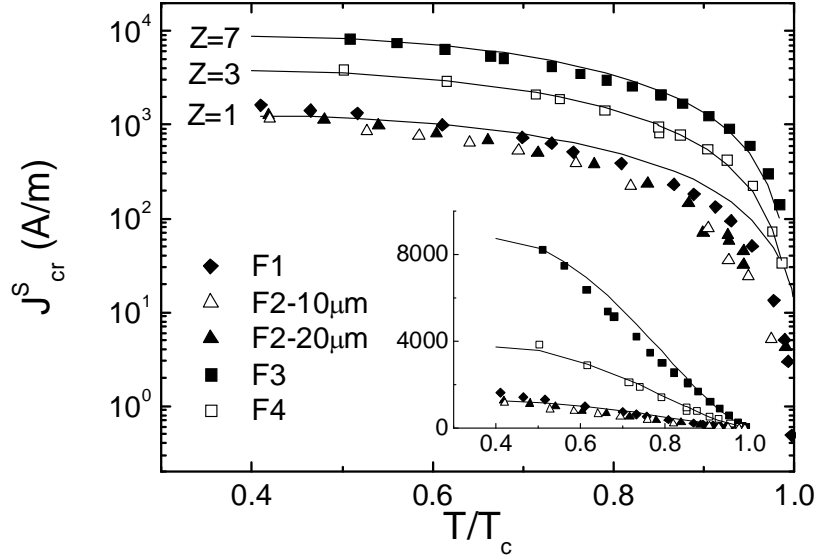

Fig.6 : Critical surface current density $J_{cr}^S = J_{cr}d$ as calculated from the $J_{cr}$ values reported in Fig.2 for the micro-bridges patterned from films F1 to F4. The solid lines show the values obtained with Eq.(8) and the Z values in Table I. The inset shows the same quantity on a linear scale.

Fig.6 compares, for the investigated micro-bridges, the experimental $J_{cr}^S\left(\dfrac{T}{T_c}\right)$ values to those computed with Eq.(8) and the Z values of Table I. There is a very good agreement except in the near vicinity of $T_c$ where Eq.(1) does not hold true for all the weak links. The corresponding $\bar{d}_s$ values are reported in Table I. We stress out that no free parameter is introduced for the calculation of $J_{cr}^S\left(\dfrac{T}{T_c}\right)$. The only parameter that depends on the considered sample is Z, that is determined from the dependence on temperature of parameter $I_o$. From a phenomenological point of view, parameter $I_o$ accounts for the curvature of the CVCs and is linked to the work carried out by the current when a vortex line section moves over distance $\bar{\delta}$ (see Eq.4b).

## 5. Domain of validity of the results

Eq.(1) is not valid above $T_{up}$. In addition, Eqs.(1), (6) and (8) are meaningless in the temperature range where $\bar{\delta}(T/T_c)$ is lower than $\xi_{ab}(T/T_c)$. A general comparison between $\bar{\delta}(T/T_c)$ and $\xi_{ab}(T/T_c)$, as given by the Ginzburg-Landau expression, is not straightforward since the relevant critical temperatures for $\bar{\delta}$ and $\xi_{ab}$ are respectively the offset and the onset temperatures. The difference between these temperatures is sample dependent. Nevertheless, the approximate value of $T_{low}$, the lower limit of the range of validity of Eq.(6), can be estimated as $\frac{T_{low}}{T_c} \approx 0.47$, taking $\xi_{ab}(T/T_c) \approx \xi_{ab}(0) = 1.41$ nm.

## C. Comparison with other measurements

Some other authors have reported $J_{cr}(T)$ values measured on YBCO films with no or a low applied field. The CVCs of the films are generally not available and it is not possible to determine Z from the comparison between Eq.(6) and $\bar{\delta}_{exp}(T/T_c)$. However, it is possible to verify if taking a suitable integer for Z, i) Eq.(1) fits $I_J(T)$ as calculated with Eq.(7c) and ii) the experimental critical surface current density of these films can be reproduced with Eq.(8).

Gonzàlez *et al.* [23] have carried out transport measurements on two sputtered YBCO micro-bridges deposited on $SrTiO_3$. The bridges were 10 µm wide and 150nm and 190nm thick, respectively. The CVCs were made available by the authors and some results obtained from these films were published in Ref.[14], especially the comparison between the quantity $J_{cr}d\bar{\delta}_{exp}$ and $I_J(T)$ as given by Eq.(1). The inset in Fig.7 compares for both films $\bar{\delta}_{exp}(T/T_c)$, as computed from the CVCs, to $\xi_{ab}(T/T_c)$. As for films F3 and F4, $\bar{\delta}_{exp}(T/T_c)$ is larger than $\xi_{ab}(T/T_c)$ in a narrow temperature range only. The $\bar{\delta}(T/T_c)$ values obtained taking Z=11 and Z=8 for the 190nm and the 150nm films, respectively and those computed with Eq.(6) are reported in the main part of Fig.7. For both films Eq.(6) fits very well the $\bar{\delta}(T/T_c)$ values in the whole range of the measurements. The critical surface current density of the films is compared in Fig.8 to the values computed with Eq.(8) taking for Z the values determined above.

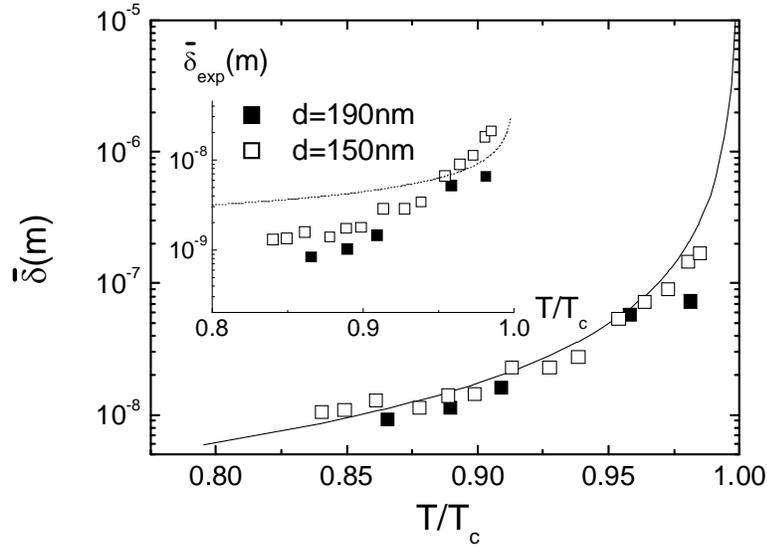

Fig.7 : Mean weak links length $\bar{\delta}(T/T_c)$ of the micro-bridges measured by Gonzàlez *et al.*[23] assuming Z=8 and Z=11 for the 150nm and the 190nm thick films, respectively. The solid line shows $\bar{\delta}(T/T_c)$ as obtained with Eq.(6). The $\bar{\delta}_{exp}(T/T_c)$ values, as computed from the CVCs, are reported in the inset. The dotted line in the inset shows $\xi_{ab}(T/T_c)$.

A very good agreement between experimental and predicted values is found for both films. The $I_J$ values computed with $\bar{\delta}$, Z and $J_{cr}$, that are reported in Fig.9, show a better agreement with Eq.(1) than those computed with $\bar{\delta}_{exp}$ and reported in Ref.[14]. The corresponding $\bar{d}_s$ values are 18.8nm for the 150nm film and 17.3nm for the 190nm film.

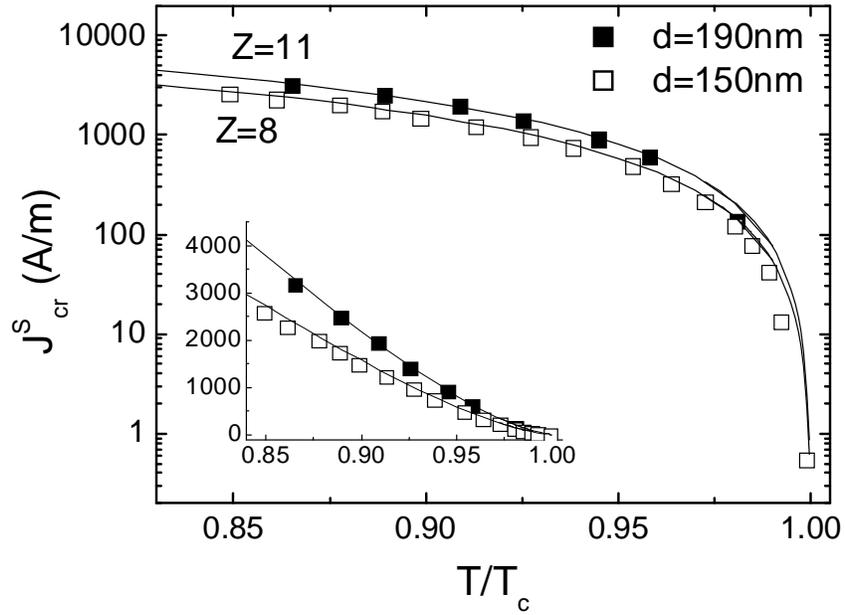

Fig.8 : Critical surface current density $J_{cr}^S = J_{cr}d$ as calculated from the $J_{cr}$ values reported by Gonzàlez *et al.*[23]. The solid lines are computed with Eq.(8) and Z=8 and Z=11, respectively for the 150nm and the 190nm films. The inset shows the same quantity on a linear scale.

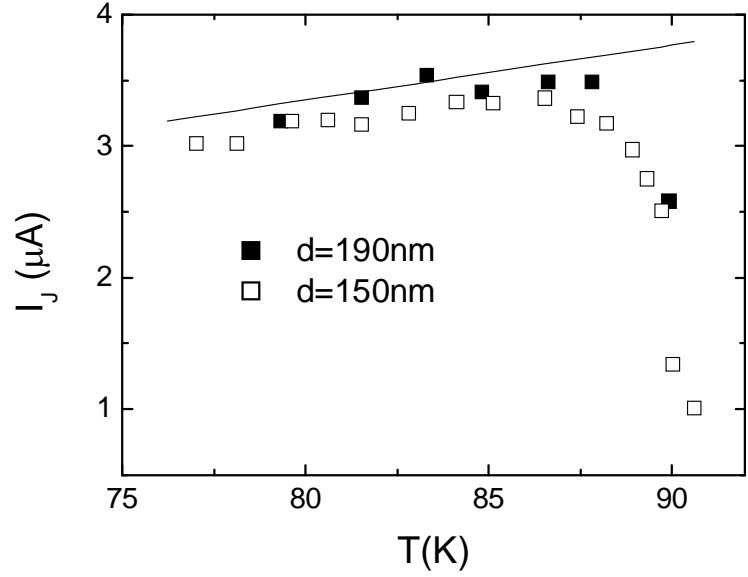

Fig.9 : Current $I_J$ flowing across a weak link of the micro-bridges measured by Gonzàlez *et al.*[23] in the critical state as a function of the temperature. The $I_J$ values are computed with Eq.(7c) from the $J_{cr}$ values reported by the authors, taking $\bar{\delta}$ as given by Eq.(6) and Z=8 and Z=11, respectively for the 150nm and the 190nm films. The solid line shows $I_J(T)$ as computed with Eq. (1).

The model can also account for the transport properties of YBCO films deposited on substrates different from $SrTiO_3$. Van der Beek *et al.*[24] have determined $J_{cr}(T)$ for YBCO films with various thicknesses that were laser ablated on $LaAlO_3$ substrates. The critical current density was estimated from magnetic and magneto-optical measurements. Fig.10 shows the critical surface current density of i) a 100nm and ii) a 250nm film. This quantity can be fairly well reproduced in the range $\frac{T}{T_c} > 0.45$ with Eq.(8) and Z=12 for the 100nm film and Z=35 for the 250nm film. The corresponding $\bar{d}_s$ values are 8.3nm and 7.1nm, respectively. The $I_J$ values are close to those calculated with Eq.(1) in the range 40K<T<80K.

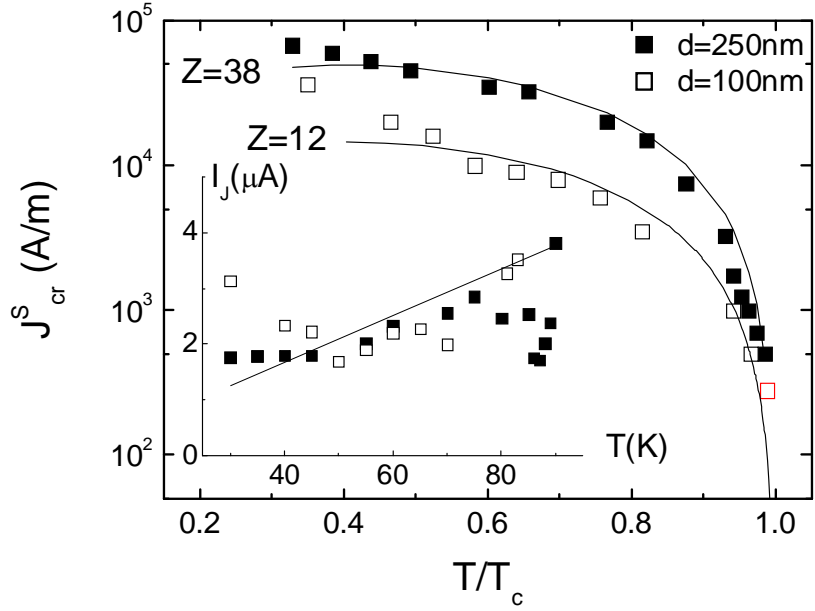

<u>Fig.10</u> : Critical surface current density $J_{cr}^{S} = J_{cr}d$ as calculated from the $J_{cr}$ values reported by van der Beek *et al.*[24] for a 100nm and a 250nm thick film. The solid lines are computed with Eq.(8) and Z=12 and Z=35, respectively. The inset shows for both films the $I_J(T)$ values computed with Eqs.(6) and (7c). The solid line reports the values given by Eq.(1).

Hudner *et al.*[25] measured with different methods $J_{cr}(T)$ for YBCO films deposited on MgO and LaAlO$_3$ substrates by different techniques. The transport measurements were carried out either on 30μmx200μm or on 10μmx200μm micro-bridges. The $J_{cr}^{S}\left(\dfrac{T}{T_c}\right)$ and $I_J(T)$ values of

i) a 280nm and a 150nm thick films, respectively co-evaporated and deposited by MOCVD on a

LaAlO$_3$ substrate and ii) a 210nm film sputtered on a MgO substrate are reported in Fig.11. Both quantities are fairly well reproduced with Eqs.(8) and (1), taking Z=8, Z=7 and Z=5 for the co-evaporated, the sputtered and the MOCVD films, respectively. The corresponding $\bar{d}_s$ values are, respectively, 30nm for both the MOCVD film deposited on LaAlO$_3$ and the film sputtered on MgO and 35nm for the co-evaporated film.

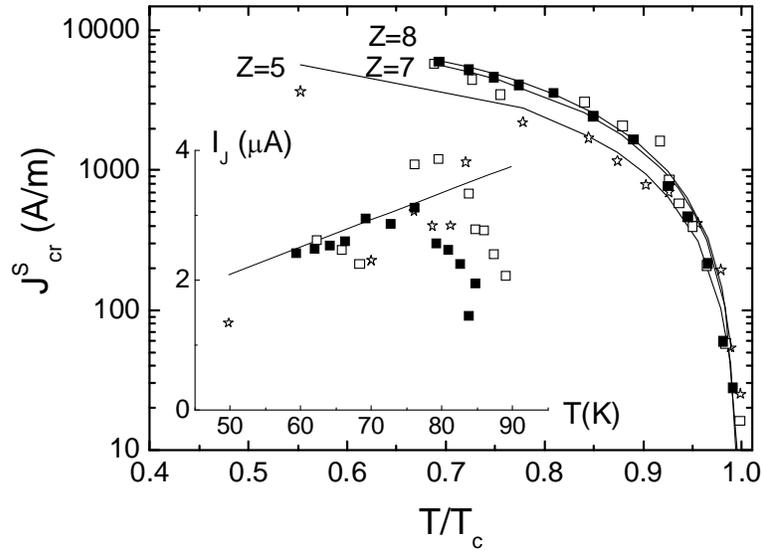

Fig.11 : Critical surface current density $J_{cr}^S = J_{cr}d$ as calculated from the $J_{cr}$ values reported by Hudner *et al.* [25] for i) a 280nm co-evaporated film on a LaAlO$_3$ substrate (full squares), ii) a 150nm film deposited by MOCVD on a LaAlO$_3$ substrate (stars) and iii) a 210nm film sputtered on a MgO substrate (open squares). The solid lines are computed with Eq.(8) and Z=8, Z=5 and Z=7, respectively. The inset shows for the three films $I_J(T)$ as computed with Eqs.(6) and (7c). The solid line reports the values given by Eq.(1).

In some cases, Eqs.(1) and (8) clearly do not account for the transport properties of the films, as shown in Fig.12 that reports $J_{cr}^S\left(\frac{T}{T_c}\right)$ and $I_J(T)$ for a 30μm×200μm YBCO bridge patterned by Hudner *et al.*[25] from a 270 nm thick film deposited by laser ablation on a MgO substrate.

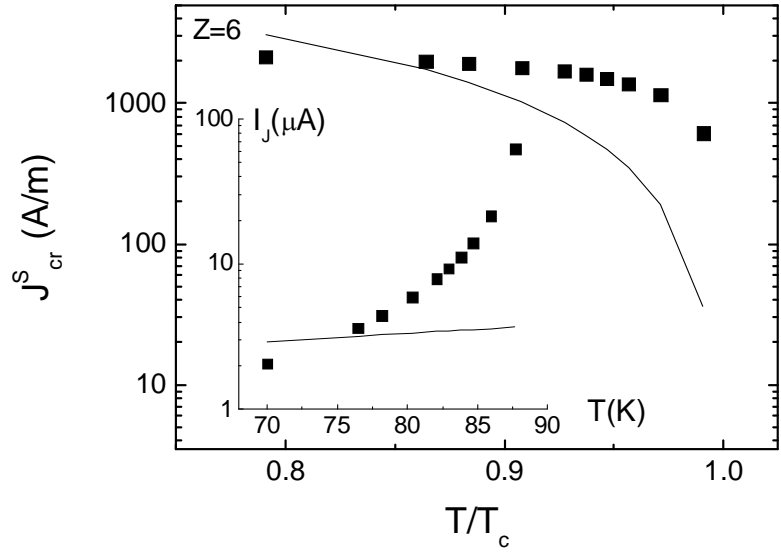

Fig.12 : Critical surface current density $J_{cr}^S = J_{cr}d$ as calculated from the $J_{cr}$ values reported by Hudner *et al.* [25] for a 270nm film laser-ablated on a MgO substrate. The solid line reports the values computed with Eq.(8) and Z=6. The inset compares the corresponding $I_J(T)$ values to those given by Eq.(1) (solid line).

**IV. Discussion**

In this section we first compare the relevance of our proposition, that the zero field vortex dynamics in twinned YBCO is that of the vortices located on the twin boundaries, to that of other propositions. Then, we discuss the contribution of the substrate and of the growth technique to the value of $\bar{d}_s$. Eventually, we consider the possible origins of the existence of the superimposed rows of weak links along the TBs.

**1. Do the twin boundaries govern the vortex dynamics in YBCO films ?**

In this section we compare the relevance of our description to that of other models. Gonzàlez *et al.*[23] have compared the CVCs they have measured to models resulting in power laws for U(I) (see section III-A-3). The main difficulty encountered by Gonzàlez *et al.*, is that the value determined for $\mu$ when fitting the CVCs is dependent on the values assigned to the other fitting parameters. As a result, they could not discriminate between the models. This difficulty was also pointed out by Schalk *et al.* [26], that obtained results very similar for U(I=0) when fitting the CVCs of their films with either $\mu=1$ or with $\mu=3/2$. Our model is limited to the physics of YBCO films in the critical state with no applied magnetic field. If an external field is applied we expect that the model is valid as far as the vortex dynamics is dominated by the motion of vortices along the TBs and if vortex interactions with the vortices either located on the same TB or nucleated in the bulk of the film can be neglected. If these conditions are not fulfilled, the model fails. Most of the measurements on YBCO films reported in the literature were carried out with a large applied field and most of the models propose a description of the vortex dynamics in the presence of a strong applied field (for an example, see Ref.[27]). Among the studies that report results obtained with a field amplitude low enough to be compared to ours, the work of Van der Beek *et al*. [24] has shown that, in these conditions, pinning in YBCO films can be ascribed neither to collective pinning, nor to variations in the films thickness, nor to the presence of screw dislocations. The authors claim that the vortices are pinned by $Y_2O_3$ inclusions (we arrive also to this conclusion that $Y_2O_3$ inclusions play an important role in the physics of the films, see section IV-3 below). To account for the critical current densities of the films that are reported in section III-C, they use as fitting parameters the density of these defects, their dimensions and the zero temperature elementary pinning force. The taken values are not unrealistic, but to fit their measurements the authors must assign different defects densities to the different films ($10^{21} m^{-3}$ and $3.10^{21} m^{-3}$ for the 100nm and 250nm thick films, respectively). For comparison, although Z was taken as a fitting parameter for the calculation of $J_{cr}^S \left( \frac{T}{T_c} \right)$ in Fig.10, the chosen values have yielded not very different $\bar{d}_s = \frac{d}{Z}$ values ($\bar{d}_s = 7.1 nm$ and $\bar{d}_s = 8.3 nm$ for the 100nm and 250nm films, respectively). Then, we consider that the proposition that the zero field vortex dynamics in twinned YBCO films is that of the vortices located on the TBs yields results that do not compare unfavorably to those yielded by the other models.

**2. Contribution of the substrate and of the growth technique to the values of $J_{cr}$ and $d_s$**

The critical surface current density of the films, as given by Eq.(8), does not depend directly on the thickness of the YBCO films but on the number of superimposed weak links rows along the TBs. This implies that the critical current density depends on $\bar{d}_s$. This quantity takes the form

$$J_{cr} = \frac{1}{\bar{d}_s} \frac{2\pi k_B T}{\phi_o \delta_o} \left(1 - \frac{T}{T_c}\right)^{\frac{3}{2}} \qquad (9)$$

in the domain of temperature where Eq.(8) is valid. From the results reported in section III, $\bar{d}_s$ depends both on the substrate and the growth technique. We however stress out that the $\bar{d}_s$ values determined for the films deposited in the same laboratory with the same technique on the same substrate, decrease as the film thickness increases. This behavior is consistent with that of the critical current density in films with $d<2\lambda_{ab}$ [24, 28]. The $\bar{d}_s$ values for films F3 and F4 that are laser ablated on SrTiO$_3$ substrates and include more than a single weak links row are 5.7 and 6.7 nm, respectively. The $\bar{d}_s$ values of the films deposited by sputtering on the same substrate by Gonzàlez *et al.* are well above this range ($\bar{d}_s \approx 18$nm). The sensitivity to the growth technique is also very strong for the films deposited on LaAlO$_3$ and MgO. The $\bar{d}_s$ values of the films deposited by laser ablation on LaAlO$_3$ substrates by Van der Beek *et al.* are in the 7.1-8.3nm range, while $\bar{d}_s$ in the cases of the films deposited by MOCVD and co-evaporation by Hudner *et al.* take values in the 30-35 nm range. This aspect requires additional work for confirmation, but the films deposited by laser ablation seem to exhibit the lowest $\bar{d}_s$ values (and the highest $J_{cr}$). In the case of the films deposited on MgO, the film deposited by sputtering shows transport properties consistent with the model, while the laser ablated film does not. This suggests that the vortex dynamics in this film is not at first determined by twinning and vortex pinning at the TBs intersections. This could be due to the 45° misorientation domains that were observed in films deposited on MgO [29] or to the treatment applied to the substrates surface before deposition, that is known to have a strong effect on their structure properties [30].

**3. Possible origins of the existence of superimposed weak links rows in YBCO films**

Contour *et al*. [31,32] have shown that above d≈10nm stress relaxation occurs during the growth of YBCO films. The result is that the part of the film located in the vicinity of the substrate is strained while the rest of the sample has relaxed. This could be the origin of two superimposed rows of weak links but can't account for the Z values larger than 2 reported in section III. Otherwise, it is known from HREM observations that there are $Y_2O_3$ inclusions in YBCO films [33,34]. The existence of this type of inclusions has been linked to a high critical current density [35]. Their density lies in the $10^{22} - 10^{24}$m$^{-3}$ range according to the substrate and the deposition technique, that corresponds to a mean distance between inclusions $d_{incl} \approx 10 - 50$nm. In some cases, the precipitates can go across a twin boundary. In any cases, although they are known to be coherent with the matrix, their presence induces strain and defects such as stacking faults, shearing planes, dislocations and twin boundaries. According to Catana *et al.* [33], a precipitate can induce extended defects parallel to the a-b planes whose length is larger than the mean distance between TBs, as sketched in Fig.13. The thickness of the resulting disordered areas is in the range of the TBs thickness. Then, it is reasonable to assume that $Y_2O_3$ inclusions cause structural disruptions in the twin boundaries. This could be the reason for the existence of superimposed weak links rows, although for the films deposited on $SrTiO_3$ by laser ablation, $\overline{d}_s$ is appreciably smaller than the observed $d_{incl}$ values.

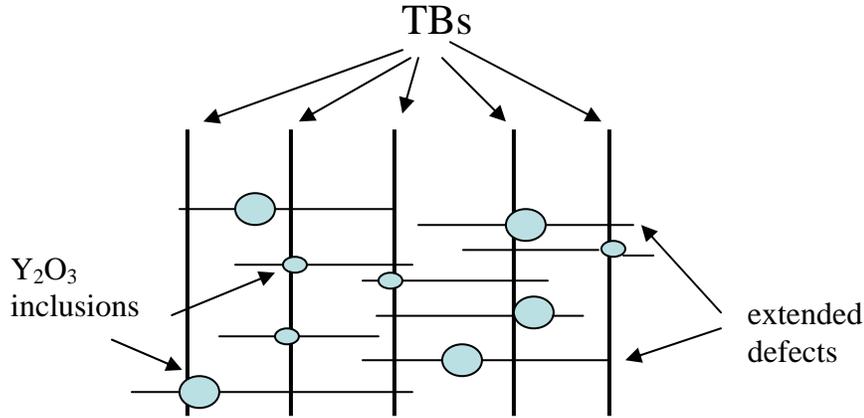

Fig.13 : Schematic representation of the structural defects induced by $Y_2O_3$ inclusions in YBCO films. The inclusions generate extended defects, such as shearing planes that can go across the TBs and create superimposed weak links rows. The grey objects represent the $Y_2O_3$ inclusions, the thick and thin lines the TBs and the extended defects, respectively.

**V. Conclusions**

In this work, we have proposed verifications and developments of the model we had previously elaborated for the superconducting transport properties of twinned YBCO films when no magnetic field is applied [13,14]. Assuming that the vortices move along the TBs of the films and that, due to the existence of disordered areas, the TBs split into rows of weak links, we have experimentally verified that the weak link energy in the critical state is equal to $k_BT$. We have shown that the supposition that films with a thickness larger than d≈10nm include superimposed rows of weak links along the TBs yields an universal expression for the mean weak links length [Eq.(6)] and another for the critical surface current density [Eq.(8)]. It is interesting that the model does not fit the experimental results in the case of some samples. This shows that the vortex dynamics is not governed by the motion of the vortices along the TBs or that the TBs intersections are not the main cause of vortex pinning in these films. For example, we don't expect that the model is valid for films with width $w \leq 2\lambda_{ab}$ that show very high critical current densities [36,37], if their transport properties depend at first on large edge barriers opposed to vortex entry, as proposed in Ref.[37]. We emphasize however, that the conclusions of this work would be the same for the most part if we had considered that the vortices are in motion along low angle grain boundaries instead of the TBs.

The existence of superimposed weak links rows could explain the enhancement of $J_{cr}(T)$ induced by the presence of $Y_2O_3$ inclusions [35]. Interestingly, a similar behavior was observed in RABiTS films, that suggests that their transport properties could also be described with the model [38].

## VI. Acknowledgments

The authors thank Maryvonne Hervieu for the help she brought them. They are grateful to M.T. Gonzàlez and her co-workers for having made available their measurements and to Kees van der Beek for his useful remarks and criticisms.